\batchmode
\documentclass[twocolumn,10pt]{asme2ej}

\usepackage{epsfig} %% for loading postscript figures
\usepackage{amsmath} %% for loading amsmath symbles
\usepackage{amssymb}
\usepackage{bm}  %% use to creat bold italic font
\usepackage{enumerate} %% creat numerate
\usepackage{graphicx} %% insert graphics
\usepackage[tight,footnotesize]{subfigure}  %% insert subgraphics
\usepackage{color}
\definecolor{MyGray}{rgb}{0.5, 0.5, 0.5}

\title{Forward Kinematics Analysis and Tension Distribution of a Cable-Driven Sinking Winches Mechanism}

\author{Xingguo Shao
\affiliation{School of Mechanical and \\Electrical Engineering\\
China University of Mining \\and Technology\\
Xuzhou, Jiangsu, China 221008\\
Email: sxgcumt@126.com}}

\author{Qingguo Wang \\
\affiliation{ Department of Electrical \& \\Computer Engineering\\
National University of Singapore\\
Email: elewqg@nus.edu.sg}}

\author{Peter C Y Chen \\
\affiliation{Department of Mechanical Engineering \\
National University of Singapore\\
Email: mpechenp@nus.edu.sg}}

\author{Zhencai Zhu \\
\affiliation{ School of Mechanical and \\Electrical Engineering\\
China University of Mining and Technology\\
Email: zhuzhencai@vip.163.com}}
\author{Bin Zi \\
\affiliation{ School of Mechanical and Electrical Engineering\\
China University of Mining and Technology\\
Email: binzicumt@163.com}}

\begin{document}

\maketitle

%%%%%%%%%%%%%%%%%%%%%%%%%%%%%%%%%%%%%%%%%%%%%%%%%%%%%%%%%%%%%%%%%%%%%%
\begin{abstract}
{\it This paper concerns the forward kinematics and tension
distribution of sinking winches mechanism, which is a type of
four-cable-driven partly constrained parallel robot. Conventional studies on forward kinematics of cable-driven parallel robot assumed that all cables are taut. Actually, given  the lengths of four cables, some cables may be slack when the platform is in static equilibrium. Therefore, in this paper, the tension state (tautness or slackness) of cables is considered in the forward kinematics model. We propose Traversal-Solving-Algorithm, which can indicate the tension state of cables, and further determine the pose of the platform, if the lengths of four cables are given. The effectiveness of the algorithm is verified by four examples. The results of this paper can be used to control sinking winches mechanism to achieve the level and stable motion of the platform, and to make the tension distribution of cables as uniform as possible.}
\end{abstract}

%%%%%%%%%%%%%%%%%%%%%%%%%%%%%%%%%%%%%%%%%%%%%%%%%%%%%%%%%%%%%%%%%%%%%%
\section{Introduction}
\begin{figure}
\centerline{\psfig{figure=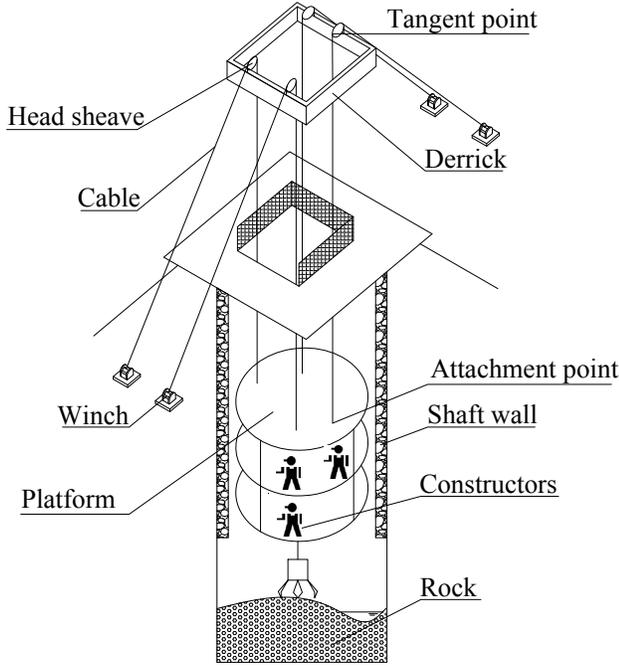,width=3.25in}}
\caption{Sinking winches mechanism}
\label{fig:swm}
\end{figure}

\noindent
In the coal mining industry, constructing a vertical shaft to transport miners and equipment down to a mine (and also to lift coal out of it) is the fist step to mine the coal. The main equipment to construct a vertical shaft is a sinking winches mechanism, as shown in Fig.~\ref{fig:swm}, which consists of four winches, four cables, a derrick and a platform. The platform, supplying the workspace for constructors, needs to be lowed down in digging process (or lifted up when cast the wall of the shaft) by rolling out (or rolling in) four cables on the winches.

The lengths of four cables from the winches to the platform determines the pose of the platform; the platform will remain horizontal if these lengths are equal. It is crucial to synchronize the motion of the winches so as to keep the platform leveled horizontally during operation, since a tilted platform in motion may collide with the wall of the shaft. However, the absolutely synchronized motion of the winches is impossible, and there always
exists synchronization error (length difference of cables). To avoid the collision of the platform and shaft wall, length difference of cables must be limited. To obtain the limitation of the length difference requires the determination of the pose of the platform given the lengths of four cables. This is the forward kinematics problem that we address in this paper.

%% why research foreward kinematics?
%% given four cables length--->pose--->check whether the collision occurs--->control the
%% lengths of four cable to avoid the collision.

A sinking winches mechanism can be considered as a cable-driven partially constrained parallel robot. The forward kinematics of parallel robots with rigid active links has been studied by many researchers. In addition to the well established methods (e.g., the elimination method\cite{Innocenti2001,Lee2001,Enferadi2010},
the continuation method\cite{Liu1995,Wampler1996}, the Gr\"oebner basis method\cite{Rouillier1995,Rouillier2003} and  the interval analysis method\cite{Merlet2004}), new approaches for dealing with this problem have been developed recently. These include methods based on neural networks\cite{Dehghani2008}, differential evolution\cite{Wang2008}, and meta-heuristic search technique\cite{Chandra2009}, and the Gauss-Newton iterative method based on chaos and hyper-chaos\cite{Luo2009}.

For forward kinematics of cable-driven parallel robots, few results are available. Most studies focused on determining the pose of the end-effector of a cable-driven fully constrained parallel robot. This is generally accomplished by solving the forward kinematics equations using  numerical method (such as Newton-Raphson method \cite{Jeong1999}) and analytical method based on the tetrahedron characteristic of the mechanism \cite{Pott2008,Zhang2009,Eusebio2010}, and only one solution was solved. Using the tool of interval analysis, Merlet\cite{Merlet2008} obtained multiple solutions for the forward kinematics equations of a robot named MARIONET.

Compared to cable-driven fully constrained parallel robots, partially constrained parallel robots (usually suspending a platform as the end-effector) represent a more difficult case in forward kinematics analysis. This difficulty arises mainly from the fact that static equilibrium equations of the platform needs to be included in the forward kinematics model\cite{Ottaviano2008}. Merlet \emph{et al}\cite{Merlet2009} studied the forward kinematics problem of a four-cable-driven parallel robot, which had the similar configuration with the sinking winches mechanism described in Figure \ref{fig:swm}, and obtained multiple poses of the platform by interval analysis; however, the multiple poses were solved based on assumption that all cables are taut. This assumption is usually unpractical since, arbitrarily giving  the lengths of four cables, the platform may be suspended by one, two, three, or four cables when the platform is in static equilibrium. Therefore, the tension state (tautness and slackness) of cables must be taken into account in the forward kinematics analysis. To investigate the forward kinematics of a sinking winches mechanism in this context is the main focus of this paper.

The remainder of this paper is organized as follows. Section 2 establishes the forward kinematics model of the sinking winches mechanism. Section 3 presents our methods for solving the forward kinematics problem, while Section 4 discusses the solving procedure. Section 5 presents four examples. Section 6 summarizes the results, and discusses their implications.

%%%%%%%%%%%%%%%%%%%%%%%%%%%%%%%%%%%%%%%%%%%%%%%%%%%%%%%%%%%%%%%%%%%%%%
\section{Forward Kinematics Model} \label{sec:FKmodel}

In this section we construct a forward kinematics model of a sinking winches mechanism driven by four cables. This model addresses geometrical constraints and static equilibrium; it incorporates the tension state, but ignores the masses and the elasticity, of the cables.

%%%%%%%%%%%%%%%%%%%%%%%%%%%%%%%%%%%%%%%%%%%%%%%%%%%%%%%%%%%%%%%%%%%%%%
\subsection{Description of the Sinking Winches Mechanism}

\begin{figure}
\centerline{\psfig{figure=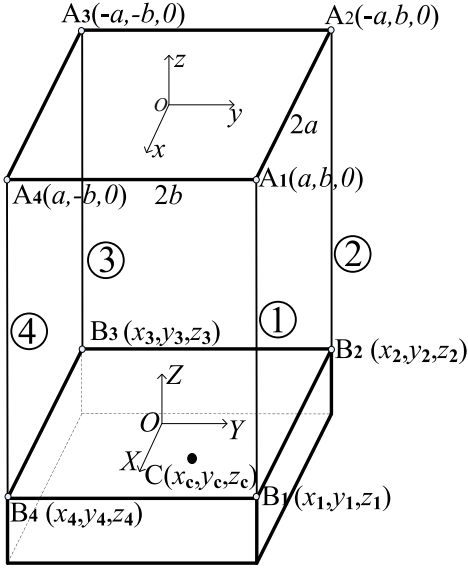,width=2.5in}}
\caption{Reduced configuration of sinking winches mechanism}
\label{fig:reducedswm}
\end{figure}

As shown in Fig.~\ref{fig:swm}, the pose of the platform is determined by the lengths of the four cables, each extending from a tangent point on the head sheave to an attachment point on the platform.
Fig.~\ref{fig:reducedswm} illustrates the relationship between the cable lengths and the pose of the platform, where $A_i$ and $B_i,(i=1,2,3,4)$ are the tangent and attachment points, respectively,  and $A_{1}A_{2}A_{3}A_{4}$\ and $B_{1}B_{2}B_{3}B_{4}$\ are identical rectangles with sides of $2a$\ and $2b$. An inertial frame $(oxyz)$\ is defined at the centroid of $A_{1}A_{2}A_{3}A_{4}$, while the local frame $(OXYZ)$\ is located at the centroid of $B_{1}B_{2}B_{3}B_{4}$. Hence, the coordinates of $B_{i}$ in the local frame are identical to those of $A_{i}$, with $i\in \{1,2,3,4\}$, in the inertial frame; these coordinates are: $(a,b,0)$, $(-a,b,0)$,$(-a,-b,0)$ and $(a,-b,0)$. In the local frame, the platform's center of gravity, denoted by $C$, is $(k_{1}a, k_{2}b,-h)$, where $k_{1}, k_{2}\in\ (-1,1)$. The geometrical parameters of the sinking winches mechanism are: $a=2$ m,  $b=2.5$ m, $h=10$ m, $k_{1}=0.25$ and $k_{2}=0.2$. The mass of the platform is $m=10^4$ kg.

The pose of the platform is represented by the Cartesian coordinates of $B_1$, $B_2$, $B_3$ and $C$ in the inertia frame. Since $B_{1}, B_{2}, B_{3}$, and $B_{4}$ are coplanar and form a rectangle, the coordinates of $B_4$ can be expressed in terms of the other three points as follows:
\begin{equation} \label{euq:corofb4}
\bm{B}_4=\bm{B}_1-\bm{B}_2+\bm{B}_3.
\end{equation}

\subsection{Geometrical Constraints}

The equations describing the geometrical constraints inherent in the sinking winches mechanism are expressed in terms of the coordinates of $B_1$, $B_2$, $B_3$ and $C$. These constraints can be classified into two groups. The first group describes the fixed distances among $B_1$, $B_2$, $B_3$ and $C$, i.e.,
\begin{equation} \label{eqn:costant distance of b1b2b3c}
\left \{ \begin{array}{l}
|\bm{B}_1\bm{B}_2|=2a \\[3pt]
|\bm{B}_2\bm{B}_3|=2b \\[3pt]
|\bm{B}_1\bm{B}_3|=2\sqrt{a^2+b^2} \\[3pt]
|\bm{C}\bm{B}_1|=r_1 \\[3pt]
|\bm{C}\bm{B}_2|=r_2 \\[3pt]
|\bm{C}\bm{B}_3|=r_3
\end{array},
\right.
\end{equation}
where, $r_1$, $r_2$ and $r_3$ denote the distances from $C$ to $B_1$, $B_2$ and $B_3$, respectively, and can be calculated as follows:
\begin{equation*}
\left \{ \begin{array}{l}
r_1=\sqrt{(1-k_1)^2a^2+(1-k_2)^2b^2+h^2} \\[3pt]
r_2=\sqrt{(1+k_1)^2a^2+(1-k_2)^2b^2+h^2} \\[3pt]
r_3=\sqrt{(1+k_1)^2a^2+(1+k_2)^2b^2+h^2}
\end{array}\,\,.
\right.
\end{equation*}

The second group of constraint equations express the lengths of taut cables in terms of the coordinates of $B_1$, $B_2$ and $B_3$. When in static equilibrium, the platform can be suspended by one, two, three, or all four cables. If the platform is suspended by a single cable, then the length of that cable can be expressed as
\begin{equation} \label{eqn:1cableconstraint}
|\bm{A}_i\bm{B}_i|=l_i \quad i \in \{1,2,3,4\}\,,
\end{equation}
where $l_{i}$ denotes the length of cable $i$.
Similarly, if the platform is suspended by two, three or four cables, then the constraint equations are
\begin{equation} \label{eqn:2cableconstraint}
\begin{array}{c}
|\bm{A}_i\bm{B}_i|=l_i, \quad |\bm{A}_j\bm{B}_j|=l_j \\
i,j \in \{1,2,3,4\},i \neq j;
\end{array}
\end{equation}
\begin{equation}\label{eqn:3cableconstraint}
\begin{array}{c}
|\bm{A}_i\bm{B}_i|=l_i, \quad |\bm{A}_j\bm{B}_j|=l_j, \quad |\bm{A}_k\bm{B}_k|=l_k \\
i,j,k \in \{1,2,3,4\},i \neq j \neq k;
\end{array}
\end{equation}
\begin{equation} \label{eqn:4cableconstraint}
|\bm{A}_i\bm{B}_i|=l_i,  \quad i=1,2,3,4.
\end{equation}
Clearly the second set equations are determined by the tension state of the cables.

%%%%%%%%%%%%%%%%%%%%%%%%%%%%%%%%%%%%%%%%%%%%%%%%%%%%%%%%%%%%%%%%%%%%%%

%%%%%%%%%%%%%%%%%%%%%%%%%%%%%%%%%%%%%%%%%%%%%%%%%%%%%%%%%%%%%%%%%%%%%%
\subsection{Static Equilibrium}
The platform is in static equilibrium under its external wrench ($\bm{F}$) and the tension of cables ($\bm{T}$). The static equilibrium can be expressed as \cite{Merlet2008}:
\begin{equation} \label{equ:staequlibrium}
{\bm{J}}^{\mathrm{T}} \cdot {\bm{T}}={\bm{F}},
\end{equation}
where ${\bm{J}}^{\mathrm{T}}$\ is the structure matrix of sinking winches mechanism, and denoted as
\begin{equation} \label{structurematrix}
{\bm{J}}^{\mathrm{T}}=\left[\begin{array}{cccc}
{\bm{u}}_1 & {\bm{u}}_2 & {\bm{u}}_3 & {\bm{u}}_4 \\
{\bm{CB}}_{1}\times{\bm{u}}_1\ & {\bm{CB}}_{2}\times{\bm{u}}_2\ & {\bm{CB}}_{3}\times{\bm{u}}_3\ & {\bm{CB}}_{4}\times{\bm{u}}_4
\end{array}
\right],
\end{equation}
where ${\bm{u}}_{i}$\ is the unit vector of ${\bm{A}}_{i}{\bm{B}}_{i}$, and calculated by ${\bm{A}}_{i}{\bm{B}}_{i}/ |{\bm{A}}_{i}{\bm{B}}_{i}|$.

Since only gravity is applied on the platform, external wrench can be expressed as
$\bm{F}=[0,0,-mg,0,0,0]^{\mathrm{T}}$. It should be emphasized that the tension vector  $\bm{T}=[{\tau}_1,{\tau}_2,{\tau}_3,{\tau}_4]^{\mathrm{T}}$, where $\tau_i$ is the tension of $i$ th cable, is determined by the tension state of cables: if the $i$ th cable is slack, $\tau_i$ must be zero.

To sum up, an example can illustrate how to construct the forward kinematics model of sinking winches mechanism. Supposing cables \textcircled{\raisebox{-1pt}{1}}, \textcircled{\raisebox{-1pt}{2}} and \textcircled{\raisebox{-1pt}{3}} are taut when the platform is in static equilibrium with the given lengths of four cables, Eqn.~(\ref{eqn:costant distance of b1b2b3c}), Eqn.~(\ref{eqn:3cableconstraint}) (with $i=1,j=2,k=3$)  and Eqn.~(\ref{equ:staequlibrium}) (with ${\bm{T}}=[{\tau}_{1},{\tau}_{2},{\tau}_{3},0]^{\mathrm{T}}$) build up the forward kimematics model. We totally obtain 15 nonlinear equations with 15 unknowns, which are tensions $({\tau}_1,{\tau}_2,{\tau}_3)$ and coordinates of $B_1$, $B_2$, $B_3$ and $C$.
Clearly, the second geometrical constraints Eqn. (\ref{eqn:3cableconstraint}) and static equilibrium Eqn. (\ref{equ:staequlibrium}) are determined by tension state of cables.

%%%%%%%%%%%%%%%%%%%%%%%%%%%%%%%%%%%%%%%%%%%%%%%%%%%%%%%%%%%%%%%%%%%%%%
\section{Solving the Forward Kinematics} \label{sec:solveFK}
\begin{figure}[!t]
\center{\subfigure[\label{subfig:aconfiguration}]{\includegraphics[width=1.5in]{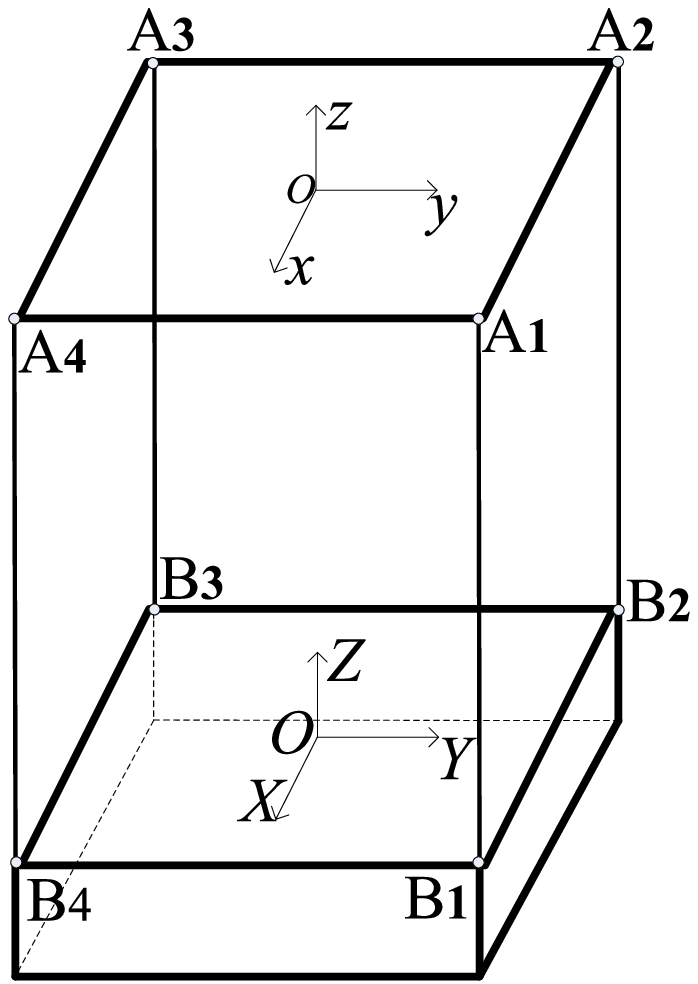}}
\subfigure[\label{subfig:bconfiguration}]{\includegraphics[width=1.6in]{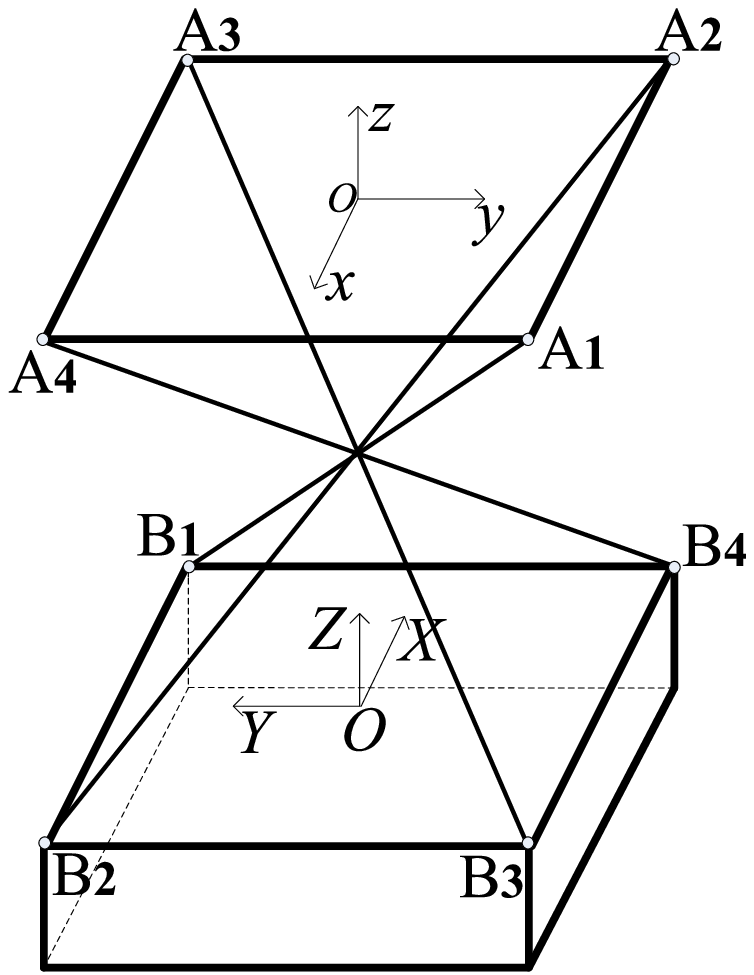}} }
\center{\subfigure[\label{subfig:cconfiguration}]{\includegraphics[width=1.5in]{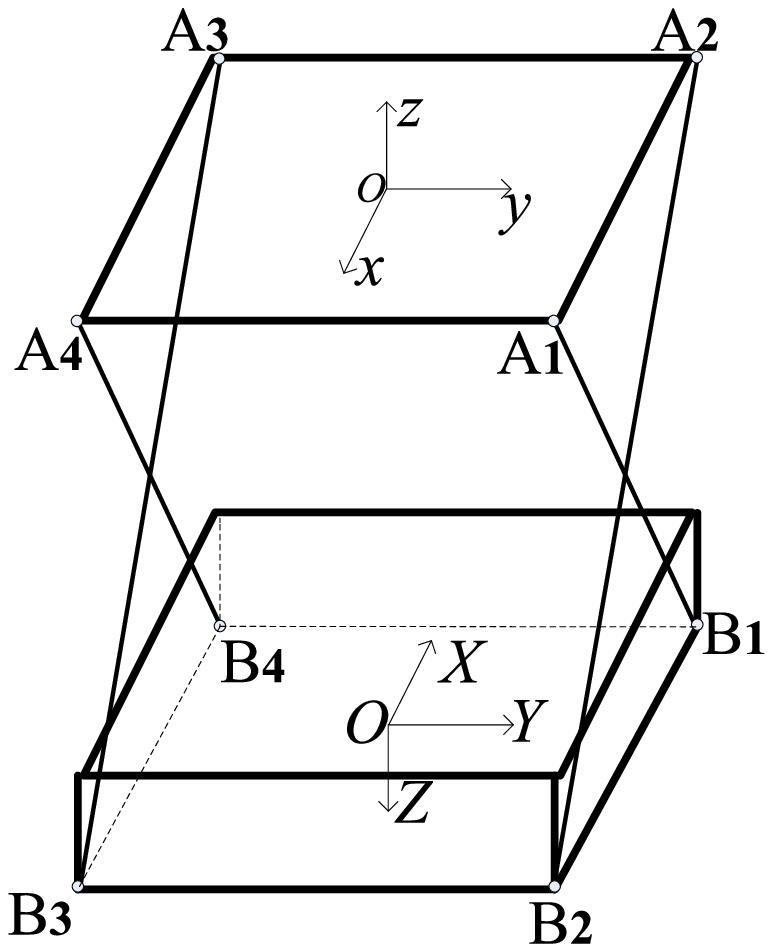}}
\subfigure[\label{subfig:dconfiguration}]{\includegraphics[width=1.5in]{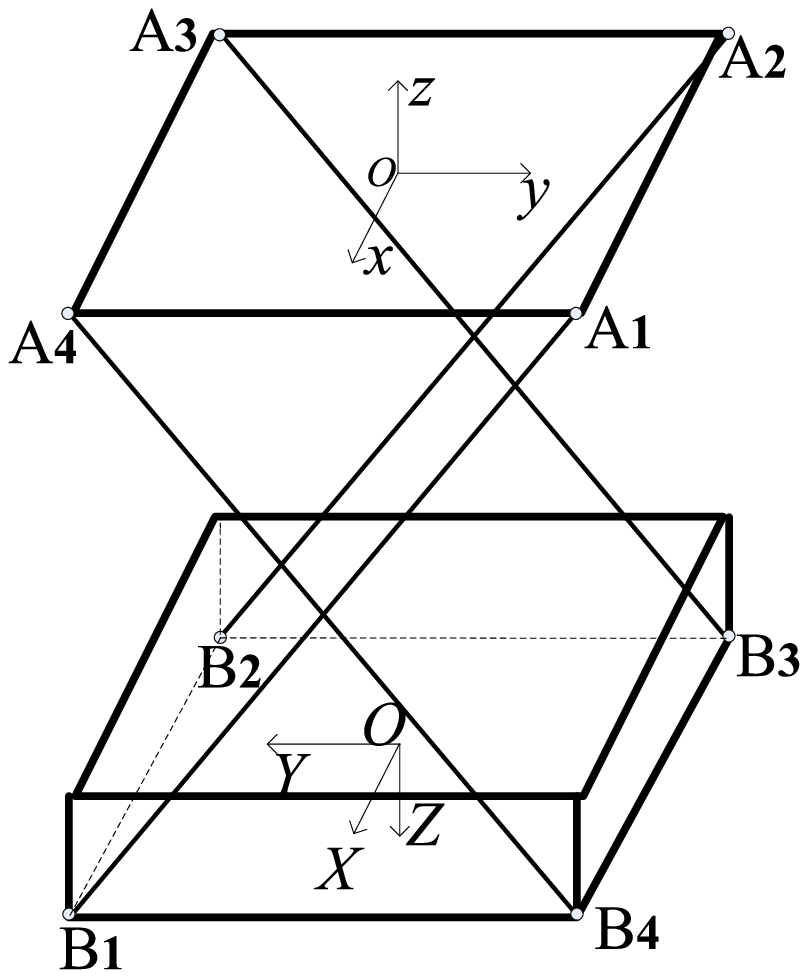}} }
\center{\subfigure[\label{subfig:econfiguration}]{\includegraphics[width=1.6in]{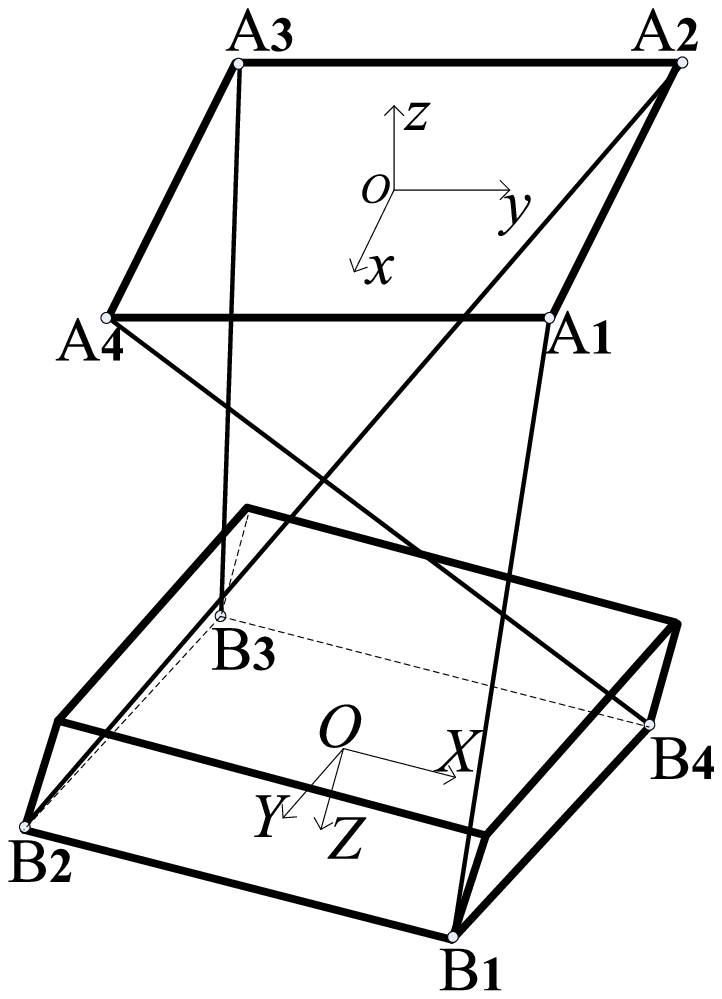}}
\subfigure[\label{subfig:fconfiguration}]{\includegraphics[width=1.4in]{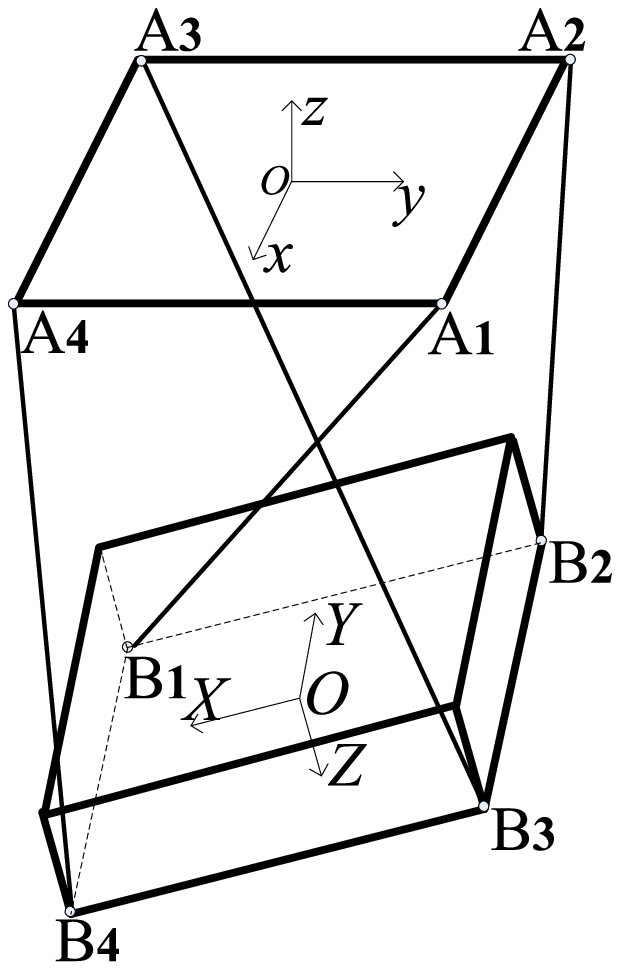}} }
\caption{Multiple configurations of SWM}
\label{fig:multipleconofswm}
\end{figure}

As described in section \ref{sec:FKmodel}, the forward kinematics model of sinking winches mechanism is constructed by a system of nonlinear equations. There exists at most six reasonable solutions for these equations since, for sinking winches mechanism, maximum six configurations (shown in Fig.~\ref{fig:multipleconofswm}) satisfy the condition that the platform is in static equilibrium.
However, this paper isn't dedicated to all reasonable solutions but focuses on the solution in the context of the first configuration (see Fig.~\ref{subfig:aconfiguration}), because it is the actual configuration of sinking winches mechanism.

It is hard to analytically solve these nonlinear equations, since the number of equations is at least 13 (if one cable is in tension). Thus, numerical method  \emph{Trust-Region Dogleg Algorithm} is selected to solve them. However, there exists three cases that cause the numerical algorithm non-convergence, or converge to wrong solutions, which are listed as follows.
\begin{enumerate}[1)]
\item Single cable suspends the platform;
\item Four cables have equal lengths;
\item Adjacent cables have equal lengths.
\end{enumerate}
The following sections give detailed analysis on reasons and propose some approaches to overcome them.

\subsection{Single Cable Suspending the Platform} \label{sec:singlecable}
If a cable is much shorter than other three cables, the platform may be solely suspended by the shortest cable when it is in static equilibrium. Without loss of generality, we assume that the platform is suspended by cable \textcircled{\raisebox{-1pt}{1}}, and cables \textcircled{\raisebox{-1pt}{2}}, \textcircled{\raisebox{-1pt}{3}}, \textcircled{\raisebox{-1pt}{4}} are slack; the tension vector is expressed as $\bm{T}=[{\tau}_1,0,0,0]^{\mathrm{T}}$. When the platform is in static equilibrium, the tangent point ($A_1$), attachment point ($B_1$), and the platform's center of gravity ($C$) form a plumb line, since external wrench applied on the platform only includes gravity. Subsequently, in inertial frame, the coordinates of $B_1$ and $C$ are $(a,b,-l_1)$ and $(a,b,-l_1-r_1)$. Thus, ${\bm{u}}_{1}={\bm{A}}_{1}{\bm{B}}_{1}/ |{\bm{A}}_{1}{\bm{B}}_{1}|=[0,0,-1]^{\mathrm{T}}$ and ${\bm{CB}}_{1} \times {\bm{u}}_1=\bm{0}$, since ${\bm{CB}}_{1}$ and ${\bm{u}}_1$ are collinear; the static equilibrium equation is expressed as
\begin{equation}\label{eqn:1cablestaequ}
{\tau}_1 \cdot [0,0,-1,0,0,0]^{\mathrm{T}}=[0,0,-mg,0,0,0]^{\mathrm{T}}.
\end{equation}
Equation (\ref{eqn:1cablestaequ}) can be reduced as ${\tau}_1=m\cdot g$, and the effective number of Eqn. (\ref{eqn:1cablestaequ}) is reduced form six to one. Combining Eqns. (\ref{eqn:costant distance of b1b2b3c}), (\ref{eqn:1cableconstraint}) and (\ref{eqn:1cablestaequ}), we end up with eight effective equations for the model of forward kinematics with thirteen unknowns. The insufficient equations for unknowns causes the ineffective and uncorrect convergence of \emph{Trust-Region Dogleg Algorithm}.

The insufficient equations for unknowns also results in multiple solutions existing for coordinates of $B_2$ and $B_3$, since tension ${\tau}_1$ and coordinates of $B_1$ and $C$ are definite. Therefore, the pose of the platform is not definite (or out of control) if the platform is suspended by single cable. This motivates us to propose the criteria to check whether this case happens with the given lengths of four cables; furthermore, we can avoid this case by control the lengths of cables.

In fact, if the platform is suspended by single cable, the other three slack cables allow the platform freely rotating around the single cable within a rotational range. Therefore, whether single cable suspends the platform can be deduced by checking if the rotational range exists. We first define critical lengths of slack cables, and then illustrate the criteria.

\subsubsection{Critical Length}
\begin{figure}
\centerline{\psfig{figure=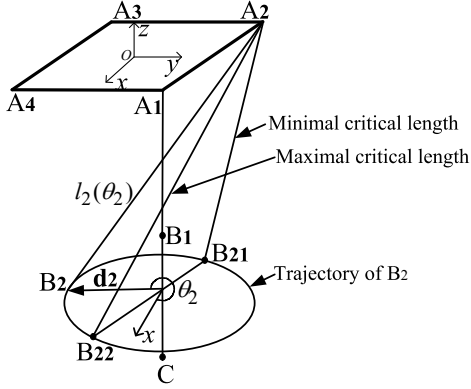,width=2.5in}}
\caption{Critical lengths of cable 2}
\label{fig:criticallengths}
\end{figure}

Assuming that cable \textcircled{\raisebox{-1pt}{1}} is shortest and solely suspend the platform, we define the critical lengths of cables \textcircled{\raisebox{-1pt}{2}}, \textcircled{\raisebox{-1pt}{3}} and \textcircled{\raisebox{-1pt}{4}} as follows.
\begin{enumerate}[1)]
  \item Rotating the platform around cable \textcircled{\raisebox{-1pt}{1}} (line $A_1B_1$) by $2\pi$, the trajectory of $B_2$ is a circle with its center locating at line $A_1B_1$, as shown in Fig.~\ref{fig:criticallengths};
  \item Arbitrarily select a point $B_2$ on the circle, let $\bm{d}_2$ be the normal vector from $B_2$ to line $A_1B_1$, and $\theta_2$ is the included angle of $\bm{d}_2$ with respect to $x$-axis of inertial frame;
  \item The distance of $A_2B_2$ is a function of $\theta_2$, and denoted as $l_2(\theta_2)=|\bm{A}_2\bm{B}_2|$;
  \item We define $l_{2\min}=\min [l_2(\theta_2)]$ and $l_{2\max}=\max [l_2(\theta_2)]$ as the minimal and maximal critical length of cable \textcircled{\raisebox{-1pt}{2}}, respectively.
\end{enumerate}
The same definitions can be made for cables \textcircled{\raisebox{-1pt}{3}} and \textcircled{\raisebox{-1pt}{4}}, which are denoted as $l_{3\min}$, $l_{3\max}$, $l_{4\min}$\ and $l_{4\max}$.

Next we deduce the values of $l_{2min}$ and $l_{2max}$. Based on geometry, the $l_{2min}$ and $l_{2max}$ are available only if $A_1B_1$ and $A_2B_2$ are coplanar. As shown in Fig.~\ref{fig:criticallengths}, let $B_{21}$, $B_{22}$ be the intersections between the circle trajectory of $B_{2}$ and the co-plane determined by $A_1B_1$ and $A_2B_2$, coordinates of $B_{21}$ and $B_{22}$ are solved by the following equations
\begin{equation*} \label{eqn:1cableGeoCons}
\left\{\begin{array}{l}
({\bm{A}}_1{\bm{B}}_1\times {\bm{B}}_1{\bm{B}}_2)\times ({\bm{A}}_2{\bm{B}}_2\times {\bm{B}}_1{\bm{B}}_2)=\mathbf{0}\\
|{\bm{B}}_1{\bm{B}}_2|=2a \\
|{\bm{C}}{\bm{B}}_2|=r_2
\end{array}.
\right.
\end{equation*}
And the coordinates of $B_{21}$ and $B_{22}$ are solved as
\begin{equation}\label{eqn:coordinateofB21}
\left\{\begin{array}{l}
x_{21}=a-\sqrt{4a^2-[(r_2^2-4a^2-r_1^2)/(2r_1)]^2} \\
y_{21}=b \\
z_{21}=(r_2^2-4a^2-r_1^2)/(2r_1) -l_1
\end{array},
\right.
\end{equation}
and
\begin{equation*}
\left\{\begin{array}{l}
x_{22}=a+\sqrt{4a^2-[(r_2^2-4a^2-r_1^2)/(2r_1)]^2} \\
y_{22}=b \\
z_{22}=(r_2^2-4a^2-r_1^2)/(2r_1) -l_1
\end{array}.
\right.
\end{equation*}
Finally, with the fixed coordinates of $A_2(-a,b,0)$, the minimal and maximal critical lengths of cable \textcircled{\raisebox{-1pt}{2}} are calculated as $l_{2min}=|\bm{A}_2\bm{B}_{21}|$ and $l_{2max}=|\bm{A}_2\bm{B}_{22}|$.
And the critical lengths of cables \textcircled{\raisebox{-1pt}{3}} and \textcircled{\raisebox{-1pt}{4}} can be solved by similar approach as cable \textcircled{\raisebox{-1pt}{2}}.

\subsubsection{Criteria} \label{criteria}
 As discussed previously, the platform is solely suspended by the shortest cable \textcircled{\raisebox{-1pt}{1}} if the lengths of cables \textcircled{\raisebox{-1pt}{2}}, \textcircled{\raisebox{-1pt}{3}} and \textcircled{\raisebox{-1pt}{4}} allow the platform freely rotating around cable \textcircled{\raisebox{-1pt}{1}} within a rotational range. This criteria can be mathematically described as:
let ${\bm{\phi}}_2$, ${\bm{\phi}}_3$ and ${\bm{\phi}}_4$ be the rotational ranges that the cables \textcircled{\raisebox{-1pt}{2}}, \textcircled{\raisebox{-1pt}{3}} and \textcircled{\raisebox{-1pt}{4}} respectively allow the platform rotating around cable \textcircled{\raisebox{-1pt}{1}}; define  ${\bm{\phi}}={\bm{\phi}}_2\bigcap{\bm{\phi}}_3\bigcap{\bm{\phi}}_4$; the platform is solely suspended by cable \textcircled{\raisebox{-1pt}{1}} if ${\bm{\phi}}$ is not empty.

From the definition of critical length, we conclude:
\begin{enumerate}[1)]
  \item If $l_j\leq l_{j\min}$, $j\in\{2,3,4\}$, the platform is not solely suspended by cable \textcircled{\raisebox{-1pt}{1}}, since cable $j$ does not allow the platform freely rotating around cable  \textcircled{\raisebox{-1pt}{1}} if the length of cable $j$ is less or equal to its minimal critical length, i.e., ${\bm{\phi}}=\emptyset$;
  \item If $l_j\geq l_{j\max}$, $j=2,3,4$, the platform is solely suspended by cable \textcircled{\raisebox{-1pt}{1}}, since the platform can rotate around cable \textcircled{\raisebox{-1pt}{1}} by $2\pi$ if the three cables (\textcircled{\raisebox{-1pt}{2}},\textcircled{\raisebox{-1pt}{3}},\textcircled{\raisebox{-1pt}{4}}) are larger or equal to their maximal critical lengths, i.e., ${\bm{\phi}}=[0,2\pi]$.
\end{enumerate}
However, what if $l_j\in (l_{j\min},l_{j\max})$, $j \in \{2,3,4\}$? In these cases, ${\bm{\phi}}_j$, $j \in \{2,3,4\}$ need to be calculated respectively to deduce whether $\bm{\phi}$ is empty. And ${\bm{\phi}}_j$ can be calculated as follows.
\begin{enumerate}[1)]
\item Suppose the platform rotating around cable \textcircled{\raisebox{-1pt}{1}} by $\theta$ from a initial pose;
\item Calculate the distances of $A_jB_j$, $j \in \{2,3,4\}$, which are expressed by $l_j(\theta)$;
\item The rotational ranges ${\bm{\phi}}_j$ are solved by the inequality $l_j(\theta) \leq l_j$ with $\theta \in [0,2\pi]$, since the distance of $A_jB_j$ is less or equal to the length of cable $j$ if $\theta \in {\bm{\phi}}_j$.
\end{enumerate}
The following gives detailed description on above approach.

\begin{figure}
\centerline{\psfig{figure=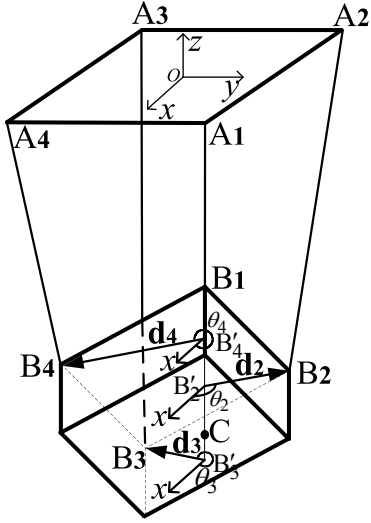,width=2in}}
\caption{Initial pose of the platform}
\label{fig:initialpose}
\end{figure}

Supposing the platform rotates around cable \textcircled{\raisebox{-1pt}{1}} by $\theta$ from the initial pose that minimal critical length of cable \textcircled{\raisebox{-1pt}{2}} is available, as shown in Fig.~\ref{fig:initialpose}, we calculate the distance $l_j(\theta)=|\bm{A}_j\bm{B}_j|$, $j \in \{2,3,4\}$. Since the coordinates of fixed points $A_j$ are known, we just need to express the coordinates of $B_j$ in term of $\theta$.
At the initial pose, we define $\bm{d}_j$, $j \in \{2,3,4\}$, the normal vector from point $B_j$ to line $A_1B_1$, whose magnitude is $d_j$, and included angel with respect to $x$-axis is ${\theta}_j$. If the platform counterclockwise rotates by $\theta$, the coordinates of $B_j$, $j \in \{2,3,4\}$, are expressed as
\begin{equation} \label{eqn:coorofBj}
\left\{\begin{array}{l}
x_j=a+d_j\cos({\theta}_j+\theta) \\
y_j=b+d_j\sin({\theta}_j+\theta) \\
z_j=z_j
\end{array},
\right.
\end{equation}
since the trajectory of $B_j$ is a circle around $A_1B_1$ in parallel with plane $xoy$; and $z_j$ keeps the same during the rotation of the platform.

In the following, we calculate $d_j$, ${\theta}_j$ and $z_j$, $j \in \{2,3,4\}$, when the platform is at the initial pose. At the initial pose, the coordinates of $B_1$ are $(a,b,-l_1)$, $B_2$ are expressed in Eqn.~(\ref{eqn:coordinateofB21}). The coordinates of $B_3$ are solved by the following equations
\begin{equation*}
\left\{\begin{array}{l}
|{\bm{B}}_1{\bm{B}}_3|=2\sqrt{a^2+b^2}\\
|{\bm{B}}_2{\bm{B}}_3|=2b \\
|{\bm{C}}{\bm{B}}_3|=r_3
\end{array}.
\right.
\end{equation*}
There exists two set of solutions for coordinates of $B_3$, and select the set of solutions satisfying the initial pose as
\begin{equation*} \label{equ:corofb3}
\left\{\begin{array}{l}
x_3=[3a^2+x_{21}^2+(z_3-z_{21})^2-(z_3+l_1)^2]/[2(x_{21}-a)] \\[3pt]
y_3=b-\sqrt{4a^2+4b^2-(z_3+l_1)^2-(x_3-a)^2} \\[3pt]
z_3= [(r_3^2-4a^2-4b^2-r_1^2)]/(2r_1)-l_1
\end{array}.
\right.
\end{equation*}
Subsequently, the coordinates of $B_4$ are calculated by substituting the coordinates
of $B_1$, $B_2$ and $B_3$ into Eqn.~(\ref{euq:corofb4}). At this point, the coordinates of $B_i$, $i \in \{1,2,3,4\}$, have been calculated, and $z_j$, $j\in \{2,3,4\}$, can be expressed as
\begin{equation*}
\left\{\begin{array}{l}
z_2=(r_2^2-4a^2-r_1^2)/(2r_1)-l_1 \\
z_3=(r_3^2-4a^2-4b^2-r_1^2)/(2r_1)-l_1 \\
z_4=(r_3^2-r_2^2-4b^2)/(2r_1)-l_1
\end{array}.
\right.
\end{equation*}
With the coordinates  of the fixed point $A_1$ and $B_i$, $i \in \{1,2,3,4\}$, $d_j$ and ${\theta}_j$, $j\in \{2,3,4\}$, can be calculated based on analytical geometry. Let $B_j'$ be the projection point of $B_j$ on line $A_1B_1$, and the coordinates of $B_j'$ in the initial frame $oxyz$ can be calculated as follows:
\begin{equation*}
\bm{B}_j'=\bm{B}_1+\bm{B}_1\bm{A}_1 \cdot \frac{\bm{B}_1\bm{A}_1 \cdot \bm{B}_1\bm{B}_j}{|\bm{B}_1\bm{A}_1|^2}, \quad j\in \{2,3,4\}.
\end{equation*}
Thus, the normal vector is expressed as $\bm{d}_j=\bm{B}_j-\bm{B}_j'$, whose magnitude $d_j$ and angles ${\theta}_j$ can subsequently calculated. Substitution of $d_j$, ${\theta}_j$, and $z_j$ into Eqn. (\ref{eqn:coorofBj}) yields coordinates of $B_j$ in term of $\theta$. Consequently, the distance $l_j(\theta)$ can be expressed by $l_j(\theta)=|\bm{A}_j\bm{B}_j|$.

With the expression of $l_j(\theta)$, $j \in \{2,3,4\}$, the rotational range $\bm{\phi}_j$ can be solved by the inequality $l_j(\theta)\leq l_j$ with $\theta \in [0,2\pi]$. To solve the inequality, we first solve the equation $l_j(\theta)=l_j$ with $\theta \in [0,2\pi]$. The equation with an interval can be solved by interval analysis\cite{Rump1999,Hansen2004}. And two solutions exist in the interval, which are denoted as $\theta_{21}$ and $\theta_{22}$ with $\theta_{21}<\theta_{22}$. Thus, the interval $[0,2\pi]$ is separated into three sub-intervals $[0,\theta_{21}]$, $[\theta_{21},\theta_{22}]$ and $[\theta_{22},2\pi]$. The solutions of the inequality  $l_j(\theta)\leq l_j$ are selected from the three sub-intervals by midpoint method, i.e., the sub-interval is the solution of the inequality if the its midpoint satisfies the inequality. And rotational range $\bm{\phi}_j$, $j\in\{2,3,4\}$ is obtained, thus, $\bm{\phi}=\bm{\phi}_2 \bigcap \bm{\phi}_3 \bigcap \bm{\phi}_4$. We can deduce whether the platform is solely suspended by cable \textcircled{\raisebox{-1pt}{1}} by checking if $\bm{\phi}$ is empty.

It should be noted that above criteria is proposed based on the assumption that cable \textcircled{\raisebox{-1pt}{1}} is shortest. For other situations, such as cable \textcircled{\raisebox{-1pt}{2}} (or \textcircled{\raisebox{-1pt}{3}}, or \textcircled{\raisebox{-1pt}{4}}) is shortest, we can take the similar analysis. Moreover, to avoid single cable suspending the platform, the lengths of cables should be controlled so as to guarantee the rotational range $\bm{\phi}$ is empty.

\subsection{Four Cables Having Equal Lengths} \label{sec:foursamecable}
\subsubsection{Pose of Platform and Tension of Cables}
For sinking winches mechanism, if four cables have equal length $l$, the platform will remain horizontal and stay at the pose that tangent point ($A_i$) and attachment point ($B_i$) of cable $i, i\in\{1,2,3,4\}$, form a plump line. At this pose, the coordinates of $B_i, i\in\{1,2,3,4\}$ and $C$ in inertial frame are $(a,b,-l)$, $(-a,b,-l)$, $(-a,-b,-l)$, $(a,-b,-l)$ and $(k_1a,k_2b,-l-h)$, respectively. At this point, the pose of the platform is obtained.

With the coordinates of $B_i, i\in\{1,2,3,4\}$ and $C$, the vectors ${\bm{u}}_i$ and ${\bm{CB}}_i$ can be calculated, which are introduced into Eqn. (\ref{structurematrix}), and the structure matrix ${\bm{J}}^{\mathrm{T}}$ is obtained. Substituting ${\bm{J}}^{\mathrm{T}}$ into Eqn. (\ref{equ:staequlibrium}), the static equilibrium of the platform is expressed as:
\begin{equation*} \label{str:maxtrix}
\underbrace{\left[
\begin{array}{cccc}
0 & 0 & 0 & 0 \\
0 & 0 & 0 & 0 \\
-1 & -1 & -1 & -1 \\
(k_2-1)b & (k_2-1)b & (k_2+1)b & (k_2+1)b \\
(1-k_1)a & (-1-k_1)a & (-1-k_1)a & (1-k_1)a \\
0 & 0 & 0 & 0
\end{array}\right]}_{{\bm{J}}^{\mathrm{T}}} \cdot \underbrace{\left[\begin{array}{c}
{\tau}_1 \\
{\tau}_2 \\
{\tau}_3 \\
{\tau}_4
\end{array}\right]}_{\bm{\tau}}=\underbrace{\left[\begin{array}{c}
0 \\
0 \\
-mg \\
0 \\
0 \\
0
\end{array}\right]}_{\bm{F}}.
\end{equation*}
There are infinite solutions for above equation, since the ranks of ${\bm{J}}^{\mathrm{T}}$ and $[{\bm{J}}^{\mathrm{T}}|{\bm{F}}]$ are both three while the number of unknowns are four. That is why \emph{Trust-Region Dogleg Algorithm} can't converge effectively and correctly. However, the above equation can be analytically solved as
\begin{equation} \label{equ:tenofcables}
\left[\begin{array}{c}
        \tau_1 \\
        \tau_2 \\
        \tau_3 \\
        \tau_4
      \end{array}
      \right]=\left[\begin{array}{c}
                      (1+k_1)mg/2 \\
                      (k_2-k_1)mg/2 \\
                      (1-k_2)mg/2 \\
                      0
                    \end{array}
      \right]+\tau_4 \cdot \left[\begin{array}{r}
                      -1 \\
                      1 \\
                      -1 \\
                      1
                    \end{array}
      \right].
\end{equation}

Considering that the tension of cables are non-negative, i.e., $\bm{\tau}\geq \bm{0}$, from which we deduce the value of ${\tau}_4$ is limited in the following intervals.
\begin{enumerate}[1)]
\item $[0,(1-k_2)mg/2]$ if $(k_1\in(-1,1),k_2\in (0,1),|k_1|\leq |k_2|)$;
\item $[0,(1+k_1)mg/2]$ if $(k_1\in(-1,0),k_2\in (-1,1),|k_1|\geq |k_2|)$;
\item $[(k_1-k_2)mg/2,(1+k_1)mg/2]$ if $(k_1\in(-1,1),k_2\in (-1,0),|k_1|\leq |k_2|)$;
\item $[(k_1-k_2)mg/2,(1-k_2)mg/2]$ if $(k_1\in(0,1),k_2\in (-1,1),|k_1|\geq |k_2|)$.
\end{enumerate}
As can be seen, the coefficients $k_1$ and $k_2$ determine the range of $\tau_4$, subsequently the tension scales of all cables, but can not determine a definite tension distribution.
Meanwhile, $k_1$ and $k_2$ also describe the location of center of gravity ($C$) on the platform, more specially, the $X$ and $Y$ coordinates of $C$ in local frame as $(k_{1}a, k_{2}b)$, $k_{1}, k_{2}\in(-1,1)$. Therefore, location of center of gravity affects the tension scales of all cables.
Among the tension scales, we are interested to the even tension distribution, i.e., tension difference among cables is minimal. This tension distribution has practical sense for the operation and maintenance of sinking winches mechanism, since it lets four cables bear even load and elastic deformation, which can extend the cables' lives and help the platform keep horizontal during the motion.

\subsubsection{Even Tension Distribution} \label{sssec:MinTenDiff}
In this section , we investigate the relation between even tension distribution and the location of center of gravity. This relation can be used to guide the designing of the platform to make tension distribution as uniform as possible.

The tension difference can be defined as
\begin{equation*}
\Delta T=\Big{[}\sum_{i\neq j}({\tau}_i-{\tau}_j)^2\Big{]}^{\frac{1}{2}}, \quad i,j \in \{1,2,3,4\}.
\end{equation*}
Substitution of Eqn.~(\ref{equ:tenofcables}) into above equation yields
\begin{equation*} \label{totaltendiff}
\begin{array}{rcl}
\Delta T & = & \big{[}16{\tau}_4^2-(8+8k_1-8k_2)mg \cdot {\tau}_4+{}    \\[2pt]
& & \displaystyle{(\frac{1+2k_1-k_2}{2}mg)^2+(\frac{k_1+k_2}{2}mg)^2+}  \\[2pt]
& & \displaystyle{(\frac{2k_2-1-k_1}{2}mg)^2+(\frac{1+k_1}{2}mg)^2+}    \\[2pt]
& & \displaystyle{(\frac{1-k_2}{2}mg)^2+(\frac{k_2-k_1}{2}mg)^2}\big{]}^{\frac{1}{2}}
\end{array}.
\end{equation*}
The above expression implies that $\Delta T$ is minimal if $\tau_4={\tau}_4^{\star}=(1+k_1-k_2)mg/4$. Besides, ${\tau}_4$ must be limited in its interval to guarantee the positive tension of cables. Therefore, the minimum of $\Delta T$ is co-determined by ${\tau}_4^{\star}$ and the interval of ${\tau}_4$. Let $\tau_{4L}$ and   $\tau_{4R}$ be the left and right endpoint of the interval, respectively. From the expression of $\Delta T$, we conclude:
\begin{enumerate}[1)]
  \item $\Delta T$ is minimal with $\tau_4=\tau_{4L}$ if ${\tau}_4^{\star}\leq \tau_{4L}$;
  \item $\Delta T$ is minimal with $\tau_4={\tau}_4^{\star}$ if $\tau_{4L} \leq {\tau}_4^{\star} \leq \tau_{4R}$;
  \item $\Delta T$ is minimal with $\tau_4={\tau}_{4R}$ if ${\tau}_4^{\star} \geq \tau_{4R}$.
\end{enumerate}
For example, the sinking winches mechanism has $k_1=0.25$ and $k_2=0.2$, which determine $\tau_4$ lying in the fourth interval $[(k_1-k_2)mg/2,(1-k_2)mg/2]=[0.025mg,0.4mg]$, thus $\tau_{4L}=0.025mg$ and $\tau_{4R}=0.4mg$. ${\tau}_4^{\star}=(1+k_1-k_2)mg/4=0.2625mg$ implies $\tau_{4L} \leq {\tau}_4^{\star} \leq \tau_{4R}$, thus the minimum of $\Delta T$ is obtained with  $\tau_4={\tau}_4^{\star}=0.2625mg$. Substitution of $\tau_4$ into Eqn.~(\ref{equ:tenofcables}) yields the tension distribution with minimal difference as: $\tau_1=0.3625mg$, $\tau_2=0.2375mg$, $\tau_3=0.1375mg$ and $\tau_4=0.2625mg$.

\begin{figure}
\centerline{\psfig{figure=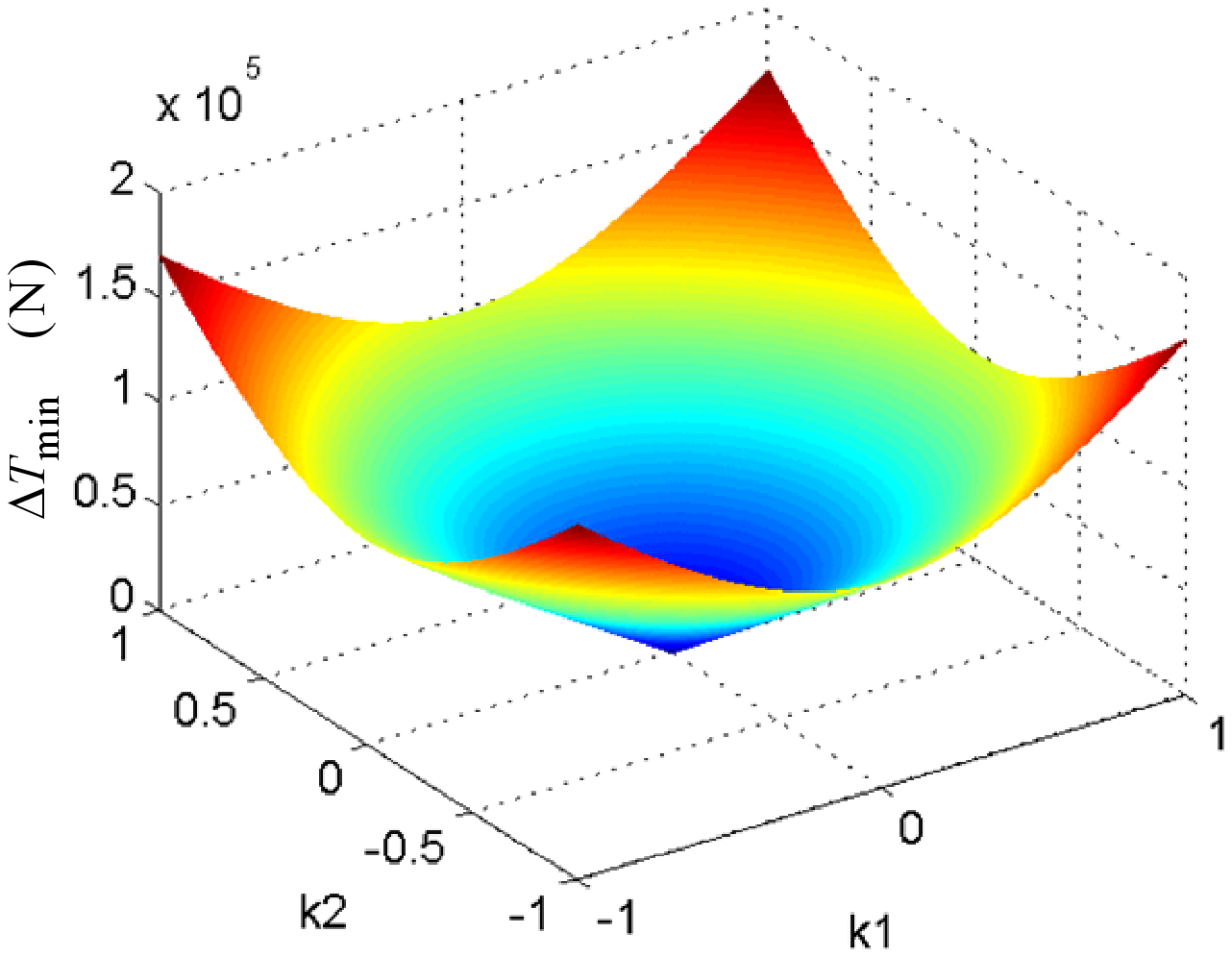,width=3.25in}}
\caption{Minimal tension difference between cables}
\label{fig:MinimalTenDiff}
\end{figure}

To illustrate the the relation of the even tension distribution and the location of center of gravity ($C$), we calculate ${\Delta T}_{min}$ with respect to $k_1,k_2 \in (-1,1)$, and the results are shown in Fig.~\ref{fig:MinimalTenDiff}. As can be seen, only if $C$ locates at the centroid of platform ($k_1=0,k_2=0$), cables can have uniform tension distribution (${\Delta T}_{min}=0$). The farther $C$ leaves from the centroid, the larger difference exists. Therefore,center of gravity should be designed at the centroid of the platform to achieve an uniform tension distribution of cables.

\subsection{Adjacent Cables Having Equal Lengths}\label{sssec:AdjCableEqual}
For sinking winches mechanism, if adjacent cables have equal lengths and all cables are taut, there exists two configurations, i.e., ($l_1=l_2$, $l_3=l_4$, $l_1 \neq l_3$) and ($l_1=l_4$, $l_2=l_3$, $l_1 \neq l_2$). For these two configurations, the structure matrix $\bm{J}^\mathrm{T}$ is nearly singular, which will be illustrated by example in section \ref{sec:example}. The nearly singularity of $\bm{J}^\mathrm{T}$ results in the ineffective and incorrect convergence of \emph{Trust-Region Dogleg Algorithm} when the algorithm is used to solve the static equilibrium equation. To overcome the defect, we firstly simplify the spacial sinking winches mechanism as a planar one because of the symmetry of its configuration; and then solve the forward kinematics of the planar mechanism; finally calculate the pose of platform and tension of cables from the forward kinematics solutions of planar mechanism.

\begin{figure}
\centerline{\psfig{figure=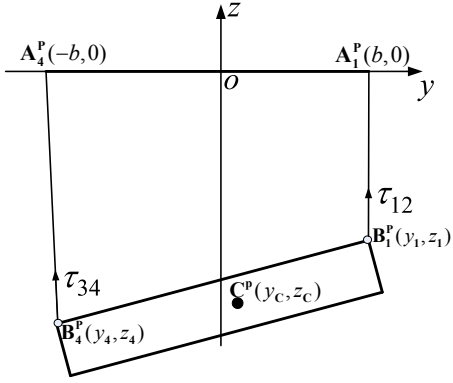,width=2.5in}}
\caption{Simplified planar configuration}
\label{fig:SimPlanarConf}
\end{figure}

Without loss of generality, taking the first configuration ($l_1=l_2$, $l_3=l_4$, $l_1 \neq l_3$) as example, the simplified planar mechanism is shown in Fig.~\ref{fig:SimPlanarConf}. This planar mechanism can be regarded as the projection of sinking winches mechanism on plane $yoz$ of inertia frame. And points $A_i$, $B_i$, $i\in\{1,2,3,4\}$, and $C$ projects at $A_1^P$, $A_4^P$, $B_1^P$, $B_4^P$, and $C^P$; besides, the projections of $B_1$ and $B_2$ are overlapped, so are $B_3$ and $B_4$. The coordinates of projection points in frame $oyz$ are shown in Fig.~\ref{fig:SimPlanarConf}, thus, the geometrical constraints are expressed as
\begin{equation} \label{equ:geoconplanar}
\left\{\begin{array}{l}
\left| \bm{A}_1^P\bm{B}_1^P\right|=l_1 \\[2pt]
\left| \bm{A}_4^P\bm{B}_4^P\right|=l_4 \\[2pt]
\left| \bm{B}_1^P\bm{B}_4^P\right|=2b  \\[2pt]
\left| \bm{C}^P\bm{B}_1^P\right|=r_1^p \\[2pt]
\left| \bm{C}^P\bm{B}_4^P\right|=r_4^p
\end{array},
\right.
\end{equation}
where $r_1^P$ and $r_4^P$ are the lengths of projections of $CB_1$ and $CB_4$ on plane $yoz$, and expressed as $r_1^P=\sqrt{(1-k_2)^2b^2+h^2}$ and $r_4^P=\sqrt{(1+k_2)^2b^2+h^2}$.

To build up the static equilibrium equation of the planar mechanism, we need to express the structure matrix ${\bm{J}}_P^{\mathrm{T}}$, the wrench applied on the platform $\bm{F}_P$, and tension of cables ${\bm{T}_P}$. ${\bm{J}}_P^{\mathrm{T}}$ are defined as:
\begin{equation*} \label{equ:jacomatrixplanar}
{\bm{J}}_P^{\mathrm{T}}=\left[\begin{array}{cc}
\displaystyle{\frac{y_1-b}{\sqrt{(y_1-b)^2+z_1^2}}} & \displaystyle{\frac{y_4+b}{\sqrt{(y_4+b)^2+z_4^2}}} \\
\displaystyle{\frac{z_1}{\sqrt{(y_1-b)^2+z_1^2}}} & \displaystyle{\frac{z_4}{\sqrt{(y_4+b)^2+z_4^2}}}\\
\displaystyle{\frac{z_1^\star}{\sqrt{(y_1-b)^2+z_1^2}}} & \displaystyle{\frac{z_4^\star}{\sqrt{(y_4+b)^2+z_4^2}}}
\end{array}
\right],
\end{equation*}
where $z_1^\star=(y_1-y_c)z_1-(y_1-b)(z_1-z_c)$ and $z_4^\star=(y_4-y_c)z_4-(y_4+b)(z_4-z_c)$; ${\bm{F}_P}$ is $[0,-mg,0]^\mathrm{T}$; ${\bm{T}_P}$ equals $[{\tau}_{12},{\tau}_{34}]^\mathrm{T}$,
where ${\tau}_{12}$ is resultant tension of cables \textcircled{\raisebox{-1pt}{1}} and \textcircled{\raisebox{-1pt}{2}}, and ${\tau}_{34}$  the resultant tension of cables \textcircled{\raisebox{-1pt}{3}} and \textcircled{\raisebox{-1pt}{4}}.
Substitution of $\bm{J}_P^\mathrm{T}$, $\bm{T}_P$, and $\bm{F}_P$ into Eqn. (\ref{equ:staequlibrium}) yields the static equilibrium equation
\begin{equation} \label{equ:staequlplanar}
\left[\begin{array}{cc}
\displaystyle{\frac{y_1-b}{\sqrt{(y_1-b)^2+z_1^2}}} & \displaystyle{\frac{y_4+b}{\sqrt{(y_4+b)^2+z_4^2}}} \\
\displaystyle{\frac{z_1}{\sqrt{(y_1-b)^2+z_1^2}}} & \displaystyle{\frac{z_4}{\sqrt{(y_4+b)^2+z_4^2}}}\\
\displaystyle{\frac{z_1^\star}{\sqrt{(y_1-b)^2+z_1^2}}} & \displaystyle{\frac{z_4^\star}{\sqrt{(y_4+b)^2+z_4^2}}}
\end{array}
\right] \cdot \left[\begin{array}{c}
{\tau}_{12} \\
{\tau}_{34}
\end{array}
\right]=\left[\begin{array}{c}
0 \\
-mg \\
0
\end{array}
\right].
\end{equation}

Equations (\ref{equ:geoconplanar}) and (\ref{equ:staequlplanar}) build up the forward kinematics model of the planar mechanism. Solving this model with \emph{Trust-Region Dogleg Algorithm} yields the coordinates of $B_1^P$, $B_4^P$ and $C^P$ in frame $oyz$, which are used to deduce the coordinates of $C$ and $B_i, i\in\{1,2,3,4\}$, in inertial frame $oxyz$. Since $B_1^P$ is the co-projection of $B_1$ and  $B_2$ on plane $yoz$, the coordinates of $B_1$ and  $B_2$ are $\bm{B}_1=(a,y_1,z_1)$ and $\bm{B}_2=(-a,y_1,z_1)$. Similarly, $B_3$ and  $B_4$ are deduced from $B_4^P$ as $\bm{B}_3=(-a,y_4,z_4)$ and $\bm{B}_4=(a,y_4,z_4)$. And $C$ is implied from $C^P$ as $\bm{C}=(k_1a,y_c,z_c)$.
With coordinates of $C$ and $B_i, i\in\{1,2,3,4\}$, the structure matrix of sinking winches mechanism ${\bm{J}}^{\mathrm{T}}$ can also be calculated through Eqn. (\ref{structurematrix}). Substituting ${\bm{J}}^{\mathrm{T}}$ into Eqn. (\ref{equ:staequlibrium}), the tension distribution of cables is calculated by the expression ${\bm{T}}=({\bm{J}}^{\mathrm{T}})^{+} \cdot {\bm{F}}$, where $({\bm{J}}^{\mathrm{T}})^{+}$ is the pseudoinverse of ${\bm{J}}^{\mathrm{T}}$. At this point, the pose of the platform and tension distribution of cables have been solved.

%%%%%%%%%%%%%%%%%%%%%%%%%%%%%%%%%%%%%%%%%%%%%%%%%%%%%%%%%%%%%%%%%%%%%%
\section{Program for Solving the Forward Kinematics} \label{sec:program}
This section present the procedure for solving the forward kinematics model (i.e., a system of nonlinear equations).
Except for the three special configurations discussed in section \ref{sec:solveFK}, forward kinematics models for other configurations can be effectively solved by \emph{Trust-Region Dogleg Algorithm}. This algorithm starts from a guess of solution to search for the solution of nonlinear equations.
As discussed in section \ref{sec:solveFK}, there exists at most six sets of reasonable solutions. Therefore, to get the unique set of solutions in the context of the first configuration (see Fig.~\ref{subfig:aconfiguration}), we need to provide a proper guess of solution for the nonlinear equations.

\subsection{Guess of Solution} \label{sec:initialguess}
The solutions of forward kinematics are the coordinates of reference points $(B_1,B_2,B_3,C)$ representing the pose of platform, and tension of cables $({\tau}_1,{\tau}_2,{\tau}_3,{\tau}_4)$.
To get the solutions in context of the first configuration, the guess of solutions are provided as follows.
\begin{enumerate}[1)]
  \item With the given lengths of four cables ($l_1,l_2,l_3,l_4$), calculate the average length as $l_{ave}=(l_1+l_2+l_3+l_4)/4$;
  \item Supposing the platform is suspended by four cables with equal length $l_{ave}$, we take the pose of platform and even tension distribution of cables as the guess of solution.
\end{enumerate}
As discussed in section \ref{sec:foursamecable}, if four cables have equal length $l_{ave}$, the coordinates of reference points are $\bm{B}_1=(a,b,-l_{ave})$, $\bm{B}_2=(-a,b,-l_{ave})$, $\bm{B}_3=(-a,-b,-l_{ave})$ and $\bm{C}=(k_1a,k_2b,-l_{ave}-h)$, and even tension distribution of cables are denoted as $({\tau}_1,{\tau}_2,{\tau}_3,{\tau}_4)_{ave}$.

Section \ref{sec:FKmodel} has elaborated that the forward kinematics model is determined by the tension state (tautness and slackness) of cables, therefore, the guess for tension need to be modified according to the tension state. For example, if cables \textcircled{\raisebox{-1pt}{1}}, \textcircled{\raisebox{-1pt}{2}} and \textcircled{\raisebox{-1pt}{3}} are taut when the platform is in static equilibrium, the guess for tension is modified as $({\tau}_1,{\tau}_2,{\tau}_3)_{ave}$.

\subsection{Traversal Solving Algorithm} \label{sec:TSA}
Section \ref{sec:singlecable} has elaborated that a definite pose of the platform can not be solved from forward kinematics model if the platform is suspend by single cable.
Except for the tension states that single cable is taut, there still exists eleven tension states : six states for two cables being taut, four states if three cables are in tension, and one case if all cables are taut. \emph{Traversal Solving Algorithm} can not only indicate the tension state of cables, but also solve the pose of platform.

\emph{Try-and-Error} is the core idea of \emph{Traversal Solving Algorithm}.
This algorithm firstly assumes a tension state; and then establish and solve the forward kinematics model corresponding to the tension state; finally check whether the assumption holds with the solutions of forward kinematics model. If the assumption holds, output the solutions; else assume another tension state, repeat above process until finding the tension state of cables.

An example can illustrate the \emph{Try-and-Error} progress: assume cables \textcircled{\raisebox{-1pt}{1}} and \textcircled{\raisebox{-1pt}{2}} being taut when the platform is in static equilibrium; Eqn. (\ref{eqn:costant distance of b1b2b3c}), Eqn. (\ref{eqn:2cableconstraint}) with $i=1,j=2$ and Eqn. (\ref{equ:staequlibrium}) with $\tau_3=\tau_4=0$ build up the forward kinematics model; solve the model by \emph{Trust-Region Dogleg Algorithm} with the guess of solutions proposed in section \ref{sec:initialguess}; calculate the distances of $A_3B_3$ and $A_4B_4$, and check whether the following inequalities hold: $|\bm{A}_3\bm{B}_3|<l_3$, $|\bm{A}_4\bm{B}_4|<l_4$, $\tau_1>0$, and $\tau_2>0$; if the inequalities hold, output the solutions, else, assume another tension state (for example, cables \textcircled{\raisebox{-1pt}{1}} and \textcircled{\raisebox{-1pt}{3}} are taut), and repeat the \emph{Try-and-Error} progress again until finding the tension state of cables.

The procedure of the \emph{Traversal Solving Algorithm} is described as follows:
\begin{enumerate}[(i)]
\item Calculate the guess of solution, go to step (ii);
\item If four cables have equal lengths, take the guess as the solution of forward kinematics, go to step (viii). Otherwise, go to step (iii);
\item Check whether the platform is suspended by single cable referring to section \ref{sec:singlecable}. If the assumption holds, output message ``The platform is suspend by single cable, an definite pose can't be solved'', go to step (viii). Otherwise, go to step (iv);
\item Suppose the platform is suspended by two cables, which include six tension states: \{\textcircled{\raisebox{-1pt}{1}},\textcircled{\raisebox{-1pt}{2}}\},
    \{\textcircled{\raisebox{-1pt}{1}},\textcircled{\raisebox{-1pt}{3}}\}, \{\textcircled{\raisebox{-1pt}{1}},\textcircled{\raisebox{-1pt}{4}}\}, \{\textcircled{\raisebox{-1pt}{2}},\textcircled{\raisebox{-1pt}{3}}\}, \{\textcircled{\raisebox{-1pt}{2}},\textcircled{\raisebox{-1pt}{4}}\}, and \{\textcircled{\raisebox{-1pt}{3}},\textcircled{\raisebox{-1pt}{4}}\}. For each tension state, establish and solve the forward kinematics model, and check whether the assumption holds. If any of the six assumption holds, go to step (viii), otherwise, go to step (v);
\item Suppose the platform is suspended by three cables, which include four tension states: \{\textcircled{\raisebox{-1pt}{1}},\textcircled{\raisebox{-1pt}{2}},\textcircled{\raisebox{-1pt}{3}}\}, \{\textcircled{\raisebox{-1pt}{1}},\textcircled{\raisebox{-1pt}{2}},\textcircled{\raisebox{-1pt}{4}}\}, \{\textcircled{\raisebox{-1pt}{1}},\textcircled{\raisebox{-1pt}{3}},\textcircled{\raisebox{-1pt}{4}}\}, and \{\textcircled{\raisebox{-1pt}{2}},\textcircled{\raisebox{-1pt}{3}},\textcircled{\raisebox{-1pt}{4}}\}. For each tension state, establish and solve the forward kinematics model, and check whether the assumption holds. If any of the four assumption holds, go to step (viii), otherwise, go to step (vi);
\item If adjacent cables have equal lengths, solve forward kinematics following the method described in section~\ref{sssec:AdjCableEqual}, go to step (viii); else, go to step (vii);
\item Solve the forward kinematics model corresponding to the tension state that four cables are taut, and go to step (viii);
\item Output the solutions, and exit the procedure.
\end{enumerate}

%%%%%%%%%%%%%%%%%%%%%%%%%%%%%%%%%%%%%%%%%%%%%%%%%%%%%%%%%%%%%%%%%%%%%%
\section{Examples} \label{sec:example}
In this section, four examples are presented to examine the \emph{Traversal Solving Algorithm}. The lengths of four cables are given in Tab. \ref{Tab.lenof4cable}.
\begin{table}[t]
\caption{Lengths of four cables}
\begin{center}
\label{Tab.lenof4cable}
\begin{tabular}{c c c c c}
& & \\ % put some space after the caption
\hline
Example & $l_1$ (m) & $l_2$ (m) & $l_3$ (m) & $l_4$ (m)\\
\hline
1 & 20 & 21 & 22 & 21.5 \\
2 & 20 & 20 & 20.1 & 20.1\\
3 & 20 & 20 & 21 & 21\\
4 & 20.3 & 20.1 & 20.5 & 20.2 \\
\hline
\end{tabular}
\end{center}
\end{table}

In example 1, the lengths of four cables are not equal, the \emph{Traversal Solving Algorithm} thus check whether the platform is solely suspended by the shortest cable \textcircled{\raisebox{-1pt}{1}} according to the criteria proposed in section \ref{criteria}. And the interval $\bm{\phi}$ is $[0,0.715]\bigcup[5.565,6.28]$ rad, where $\bm{\phi}$ is the rotational range that cables \textcircled{\raisebox{-1pt}{2}}, \textcircled{\raisebox{-1pt}{3}}, and \textcircled{\raisebox{-1pt}{4}} allow the platform freely rotating around cable \textcircled{\raisebox{-1pt}{1}}. The non-empty of $\bm{\phi}$ implies that the platform is solely suspended by cable \textcircled{\raisebox{-1pt}{1}}. Therefore, a definite pose of the platform can not be solved.
As for examples 2$\sim$4, the intervals $\bm{\phi}$ are empty. Consequently, the pose of platform and tension distribution of cables can be solved with \emph{Traversal Solving Algorithm}, and the solutions are list in Tab. \ref{Tab.posetension}.

\begin{table}[t]
\caption{Platform pose and cable tension}
\begin{center}
\label{Tab.posetension}
\begin{tabular}{l r r r}
& & \\ % put some space after the caption
\hline
Example & 2 & 3 & 4 \\
\hline
$B_{1X}$ & 2.000   & 2.000     & 1.996 \\
$B_{1Y}$ & 2.499   & 2.500     & 2.499 \\
$B_{1Z}$ & -19.999  & -20.000   & -20.299 \\
$B_{2X}$ & -2.000  & -2.000    & -1.999 \\
$B_{2Y}$ & 2.499   & 2.500     & 2.499 \\
$B_{2Z}$ & -19.999 & -20.000   & -20.099 \\
$B_{3X}$ & -2.000  & -2.000    & -1.995 \\
$B_{3Y}$ & -2.499  & -2.403    & -2.499 \\
$B_{3Z}$ & -20.099 & -20.981   & -20.000 \\
$C_{X}$  & 0.500   & 0.500     & -0.001 \\
$C_{Y}$  & 0.700   & 2.500     & 0.299 \\
$C_{Z}$  & -30.038 & -30.198   & -30.170 \\
$\tau_1$ (kN) & 39.201 & 61.250 & 5.856 \\
$\tau_2$ (kN) & 23.520 & 38.750 & 49.018 \\
$\tau_3$ (kN) & 13.229 & 0      &   0 \\
$\tau_4$ (kN) & 22.049 & 0      &  43.126 \\
\hline
\end{tabular}
\end{center}
\end{table}

In example 2, since $l_1=l_2$, $l_3=l_4$, and $l_1 \neq l_3$, adjacent cables have equal lengths. Moreover, all cables are taut, since the assumptions that two or three cables are taut do not hold. As described in section \ref{sssec:AdjCableEqual}, the forward kinematics is solved by simplifying the spacial configuration as a planar one. With the coordinates of reference points $B_1$, $B_2$, $B_3$ and $C$, the structure matrix ${\bm{J}}^{\mathrm{T}}$ is calculated as
\begin{equation*} \label{exp:strmatrixnum}
{\bm{J}}^{\mathrm{T}}=\left[
\begin{array}{cccc}
0 & 0 & 0 & 0 \\
-0.00002 & -0.00002 & 0.00003 & 0.00003 \\
-1 & -1 & -1 & -1 \\
-1.79942 & -1.79942 & 3.19908 & 3.19908 \\
1.5 & -2.5 & -2.5 & 1.5 \\
-0.00003 & 0.00004 & -0.00008 & 0.00005
\end{array}
\right].
\end{equation*}
The singular values of ${\bm{J}}^{\mathrm{T}}$ are 5.43464, 4.08118, 1.32388, 0.00008. Since there exist a singular value (0.00008) closing to 0, the ${\bm{J}}^{\mathrm{T}}$ is nearly singular, which results in the ineffective and incorrect convergence of \emph{Trust-Region Dogleg Algorithm}.

In example 3, even if $l_1=l_2$, $l_3=l_4$ and $l_1 \neq l_3$, the platform is hanged only by cables \textcircled{\raisebox{-1pt}{1}}, \textcircled{\raisebox{-1pt}{2}}, since cables \textcircled{\raisebox{-1pt}{3}}, \textcircled{\raisebox{-1pt}{4}} are much longer than \textcircled{\raisebox{-1pt}{1}}, \textcircled{\raisebox{-1pt}{2}}.
As for example 4, the platform is hanged by cables \textcircled{\raisebox{-1pt}{1}}, \textcircled{\raisebox{-1pt}{2}} and \textcircled{\raisebox{-1pt}{4}} when it is in static equilibrium.

%%%%%%%%%%%%%%%%%%%%%%%%%%%%%%%%%%%%%%%%%%%%%%%%%%%%%%%%%%%%%%%%%%%%%%
\section{Conclusions}
This paper examines the forward kinematics and tension distribution of sinking winches mechanism. The tension state of cables is considered in the forward kinematics model. And the tension state affects both geometrical constraints and statical equilibrium equations. When the platform is in static equilibrium,
there exist at most six configurations, i.e., maximum six reasonable solutions exist for the forward kinematics model. However, this paper only focus on the solution in the context of the first configuration, which agrees with the configuration of sinking winches mechanism.

\emph{Traversal Solving Algorithm} is proposed to solve the forward kinematics model. In particular, three special configurations that cause the ineffective and uncorrect convergence of the numerical algorithm are detailed discussed:
\begin{enumerate}[1)]
         \item A definite pose of the platform can not be solved from the forward kinematic model if the platform is suspended by single cable; and the criteria is proposed to check whether this case happens;
         \item If four cables' lengthes are equal, a definite tension distribution can not be obtained from the forward kinematic analysis, but the even tension distribution is unique;
         \item When adjacent cables have equal lengths and all cables are taut, the forward kinematics can be analyzed by simplifying the spacial mechanism as a planar one.
\end{enumerate}

The analysis of forward kinematics yields the pose of platform and tension distribution of cables if the lengths of four cables are given. With the pose, we can check whether the platform collide with the shaft wall, furthermore control the lengths of cables to achieve the level and stable motion of the platform. To achieve an uniform tension distribution, the platform's center of gravity must be designed at the centroid of the platform.

In this paper, the static equilibrium of platform is considered in the forward kinematics analysis. Actually, during the motion of platform, the inertial force of the platform can not be neglected, because the mass of platform take much account of the whole mechanism. Furthermore, the mass and elasticity of cable play important role in the motion of the mechanism. Therefore, the dynamics model of the mechanism is our future work.

%%%%%%%%%%%%%%%%%%%%%%%%%%%%%%%%%%%%%%%%%%%%%%%%%%%%%%%%%%%%%%%%%%%%%%
\begin{acknowledgment}
We are indebted to Prof. J.-P. Merlet for discussion on the forward kinematics of the four cables based parallel mechanism. We would like to express the gratitude to Tiago Montanher and Prof. Siegfried M. Rump for their help and suggestions on solving the nonlinear equations with interval analysis. We also appreciate the supports of National 863 Program of China (Grant Number 2009AA04Z415) and Nature Science Foundation of China (Grant Number 50905179).
\end{acknowledgment}

%%%%%%%%%%%%%%%%%%%%%%%%%%%%%%%%%%%%%%%%%%%%%%%%%%%%%%%%%%%%%%%%%%%%%%

\bibliographystyle{asmems4}
\bibliography{FKOFSWM}

\begin{thebibliography}{10}

\bibitem{asmemanual}
{ASME}, 2003.
\newblock {\em {ASME} Manual {MS-4}, An {ASME} Paper}, latest~ed.
\newblock The American Society of Mechanical Engineers, New York.
\newblock See also URL \verb+http://www.asme.org/pubs/MS4.html+.

\bibitem{latex}
Lamport, L., 1986.
\newblock {\em \LaTeX: a Document Preparation System}.
\newblock Addison-Wesley, Reading, MA.

\bibitem{goosens}
Goosens, M., Mittelbach, F., and Samarin, A., 1994.
\newblock {\em The \LaTeX\ Companion}.
\newblock Addison-Wesley, Reading, MA.

\bibitem{art}
Author, A., Author, B., and Author, C., 1994.
\newblock ``Article title''.
\newblock {\em Journal {N}ame, {\bf 1}}(5), May, pp.~1--3.

\bibitem{blt}
Booklet, A., 1994.
\newblock Booklet title.
\newblock On the WWW, at \verb+http://www.abc.edu+, May.
\newblock PDF file.

\bibitem{ibk}
Inbook, A., ed., 1991.
\newblock {\em Book title}, $1^{st}$~ed., Vol.~2 of {\em {Series Title}}.
\newblock Publisher {N}ame, Publisher address, {Chap.}~1, pp.~1--3.
\newblock See also URL \verb+http://www.abc.edu+.

\bibitem{icn}
Incollection, A., 1991.
\newblock ``Article title''.
\newblock In {\em Collection {T}itle}, A.~Editor, ed., $3^{rd}$~ed., Vol.~2 of
  {\em Series title}. Publisher {N}ame, Publisher address, May, {C}hapter~1,
  pp.~1--3.
\newblock See also URL \verb+http://www.abc.edu+.

\bibitem{ips}
Inproceedings, A., 1991.
\newblock ``Article title''.
\newblock In Proceedings {T}itle, A.~Editor and B.~Editor, eds., Vol.~{\bf 1}
  of {\em Series name}, Organization {N}ame, Publisher {N}ame, pp.~1--3.
\newblock Paper number 1234.

\bibitem{mts}
Mastersthesis, A., 2003.
\newblock ``{Thesis Title}''.
\newblock {MS Thesis}, University of Higher Education, Cambridge, {MA}, May.
\newblock See also URL \verb+http://www.abc.edu+.

\bibitem{mis}
Misc, A., 2003.
\newblock Miscellaneous {T}itle.
\newblock On the WWW, May.
\newblock URL \verb+http://www.abc.edu+.

\bibitem{pro}
Proceedings, A., ed., 1991.
\newblock Volume {T}itle, Vol.~1 of {\em Proceedings {S}eries}, Organization
  {N}ame, Publisher {N}ame.
\newblock See also URL \verb+http://www.abc.edu+.

\bibitem{pts}
Phdthesis, A., 2003.
\newblock ``{Thesis Title}''.
\newblock {PhD Thesis}, University of Higher Education, Cambridge, {MA}, May.
\newblock See also URL \verb+http://www.abc.edu+.

\bibitem{trt}
Techreport, A., 2003.
\newblock {Techreport title}.
\newblock Progress report~1, University of Higher Education, Cambridge, {MA},
  May.
\newblock See also URL \verb+http://www.abc.edu+.

\bibitem{upd}
Unpublished, A., 2003.
\newblock {Unpublished document title}.
\newblock See also URL \verb+http://www.abc.edu+, May.

\end{thebibliography}


\begin{thebibliography}{10}

\bibitem{Innocenti2001}
Innocenti, C., 2001.
\newblock ``Forward kinematics in polynomial form of the general stewart
  platform''.
\newblock {\em ASME Journal of Mechanical Design, {\bf 123}}, Jun,
  pp.~254--260.

\bibitem{Lee2001}
Lee, T.-Y., and Shim, J.-K., 2001.
\newblock ``Forward kinematics of the general 6-6 stewart platform using
  algebraic elimination''.
\newblock {\em Mechanism and Machine Theory, {\bf 36}}, pp.~1073--1085.

\bibitem{Enferadi2010}
Enferadi, J., and Tootoonchi, A.~A., 2010.
\newblock ``A novel approach for forward position analysis of a double-triangle
  spherical parallel manipulator''.
\newblock {\em European Journal of Mechanics A/Solids, {\bf 29}}, pp.~348--355.

\bibitem{Liu1995}
Liu, A.-X., and Yang, T.-L., 1995.
\newblock ``Configuration analysis of a class of parallel structures using
  improved continuation''.
\newblock In 9thWorld Congress on the Theory of Machines and Mechanisms,
  pp.~155--158.

\bibitem{Wampler1996}
Wampler, C., 1996.
\newblock ``Forward displacement analysis of general six-in-parallel sps
  (stewart) platform manipulators using soma coordinates''.
\newblock {\em Mechanism and Machine Theory, {\bf 31}}(3), pp.~331--337.

\bibitem{Rouillier1995}
Rouillier, F., 1995.
\newblock ``Real roots counting for some robotics problems''.
\newblock In Computational Kinematics, J.-P. Merlet and B.~Ravani, eds., Kluwer
  Academic Publishers, pp.~73--82.

\bibitem{Rouillier2003}
Rouillier, F., 2003.
\newblock ``Efficient real solutions and robotics''.
\newblock In First EMS-SMAI-SMF Joint Conference on Applied Mathematics and
  Applications of Mathematics.

\bibitem{Merlet2004}
Merlet, J.-P., 2004.
\newblock ``Solving the forward kinematics of a gough-type parallel manipulator
  with interval analysis''.
\newblock {\em The International Journal of Robotics Research, {\bf 3}},
  pp.~221--235.

\bibitem{Dehghani2008}
Dehghani, M., Ahmadi, M., Khayatian, A., Eghtesad, M., and Farid, M., 2008.
\newblock ``Neural network solution for forward kinematics problem of hexa
  parallel robot''.
\newblock In 2008 American Control Conference, pp.~4214--4219.

\bibitem{Wang2008}
Wang, X., Hao, M., and Cheng, Y., 2008.
\newblock ``On the use of differential evolution for forward kinematics of
  parallel manipulators''.
\newblock {\em Applied Mathematics and Computation, {\bf 205}}, pp.~760--769.

\bibitem{Chandra2009}
Chandra, R., Frean, M., and Rolland, L., 2009.
\newblock ``A meta-heuristic paradigm for solving the forward kinematics of 6-6
  general parallel manipulator''.
\newblock In IEEE International Symposium on Computational Intelligence in
  Robotics and Automation, pp.~171--176.

\bibitem{Luo2009}
Luo, Y., and Liu, Q., 2009.
\newblock ``Chaotic finding method of position forward kinematics of 3-dof
  parallel robot''.
\newblock In Third International Symposium on Intelligent Information
  Technology Application, pp.~44--47.

\bibitem{Jeong1999}
Jeong, J.~W., Kim, S.~H., and Kwak, Y.~K., 1999.
\newblock ``Kinematics and workspace analysis of a parallel wire mechanism for
  measuring a robot pose''.
\newblock {\em Mechanism and Machine Theory, {\bf 34}}, pp.~825--841.

\bibitem{Pott2008}
Pott, A., 2008.
\newblock ``Forward kinematics and workspace determination of a wire robot for
  industrial applications''.
\newblock In Advances in Robot Kinematics: Analysis and Design, pp.~451--458.

\bibitem{Zhang2009}
Zhang, J., Li, J., Qi, L., and Zhang, D., 2009.
\newblock ``Kinematic analysis of a 6-dof wire-based tracking device and
  control strategy for its application in robot easy programming''.
\newblock In Proceedings of the 2009 IEEE International Conference on Robotics
  and Biomimetics, pp.~1591--1596.

\bibitem{Eusebio2010}
Hern\'andez-Mart\'inez, E.~E., Ceccarelli, M., Carbone, G., L\'opez-Caj¨²n,
  C.~S., and J\'auregui-Correa, J.~C., 2010.
\newblock ``Characterization of a cable-based parallel mechanism for
  measurement purposes''.
\newblock {\em Mechanics Based Design of Structures and Machines, {\bf 38}},
  pp.~25--49.

\bibitem{Merlet2008}
Merlet, J.-P., 2008.
\newblock ``Kinematics of the wire-driven parallel robot marionet using linear
  actuators''.
\newblock In IEEE International Conference on Robotics and Automation.

\bibitem{Ottaviano2008}
Ottaviano, E., 2008.
\newblock ``Analysis and design of a four-cable-driven parallel manipulator for
  planar and spatial tasks''.
\newblock In Proceedings of the Institution of Mechanical Engineers, Part C:
  Journal of Mechanical Engineering Science, Vol.~222 of {\em 8},
  pp.~1583--1592.

\bibitem{Merlet2009}
Merlet, J.-P., and Daney, D., 2009.
\newblock ``Kinematic analysis of a spatial four-wire driven parallel crane
  without constraining mechanism''.
\newblock In Computational Kinematics: Proceedings of the 5th International
  Workshop on Computational Kinematics, A.~Muller and A.~Kecskemethy, eds.,
  Springer, pp.~1--8.

\bibitem{Rump1999}
Rump, S., 1999.
\newblock {\em Developments in Reliable Computing}.
\newblock Kluwer Academic Publishers.

\bibitem{Hansen2004}
Hansen, E., and Walster, G.~W., 2004.
\newblock {\em Global Optimization Using Interval Analysis}.
\newblock CRC Press, ch.~Systems of Nonlinear Equations, pp.~237--282.

\end{thebibliography}

%%%%%%%%%%%%%%%%%%%%%%%%%%%%%%%%%%%%%%%%%%%%%%%%%%%%%%%%%%%%%%%%%%%%%%

\end{document}